\documentclass[twocolumn,english]{revtex4}
\usepackage[T1]{fontenc}
\usepackage[utf8]{luainputenc}
\usepackage{amsmath}
\usepackage{amssymb}
\usepackage{graphicx}
\usepackage{esint}

\makeatletter
\@ifundefined{textcolor}{}
{%
 \definecolor{BLACK}{gray}{0}
 \definecolor{WHITE}{gray}{1}
 \definecolor{RED}{rgb}{1,0,0}
 \definecolor{GREEN}{rgb}{0,1,0}
 \definecolor{BLUE}{rgb}{0,0,1}
 \definecolor{CYAN}{cmyk}{1,0,0,0}
 \definecolor{MAGENTA}{cmyk}{0,1,0,0}
 \definecolor{YELLOW}{cmyk}{0,0,1,0}
}

\usepackage{babel}

\usepackage{babel}

\usepackage{babel}

\usepackage{babel}

\usepackage{babel}

\usepackage{babel}

\usepackage{babel}

\usepackage{babel}

\usepackage{babel}

\makeatother

\usepackage{babel}
\begin{document}

\title{Equilibration and Generalized GGE in the Lieb Liniger gas}

\author{G. Goldstein and N. Andrei}

\address{Department of Physics, Rutgers University, Piscataway, New Jersey
08854}
\begin{abstract}
We study the nonequilibrium properties of the 1-D Lieb-Liniger model
in the thermodynamic limit for \emph{finite} repulsive coupling. For
this purpose we introduce a new version of the Yudson representation
applicable to finite size systems and obtain the thermodynamic limit
by appropriately taking the infinite volume - at constant density
- limit. We provide a formalism to compute various correlation functions
for highly non-equilibrium initial states. In the strong coupling
limit we are able to find explicit analytic expressions for the expectation
of the density, density density and related correlation functions
at arbitrary times. We present our result as a power series expansion
in inverse coupling strength. We show that the gas equilibrates to
a steady state from arbitrary initial states with ``smooth'' correlation
functions. For nearly translationally invariant states the gas equilibrates
to a diagonal ensemble which we show is equivalent to a generalized
version of the GGE for sufficiently simple correlation functions,
which in particular include density density correlations. 
\end{abstract}
\maketitle
\textit{Introduction.} Non-equilibrium processes can be found in many
diverse fields ranging from biology to metallurgy to quantum chemistry.
In all these instances it is of interest to study the evolution of
a system away from thermodynamic equilibrium. In theoretical physics
non-equilibrium processes play an important role, underlying statistical
mechanics, transport theory, linear and non-linear response theory.
One of the key questions under study is the equilibration of a macroscopic
system initiated far from thermal equilibrium, see e.g. \cite{key-32,key-11}.
The study of equilibration phenomena has recently received a boost
from the field of cold atoms, where quench experiments can be carried
out in a highly controlled manner: by a judicious use of external
lasers one creates a system in well defined state $\left|\Phi\left(t=0\right)\right\rangle $.
Then through a rapid change in the parameters of the system, in the
cold atom setup this would typically correspond to an adjustment of
the external lasers and magnetic fields, the experimenter is able
to modify the dynamics of the system, create a new effective Hamiltonian
$H$ and induce time evolution: $\left|\Phi\left(t=0\right)\right\rangle \to e^{-iHt}\left|\Phi\left(t=0\right)\right\rangle $.
This way one may study interesting correlation effects and equilibration
properties. Furthermore the final state of the system, say its density
density correlation function, can be effectively measured through
time of flight experiments and absorption imaging \cite{key-17,key-7}.

Many of the systems currently under such study may be well described
by integrable models, ones with an infinite number of conserved quantities
\cite{key-1,key-16}. The eigenstates $|k\rangle$ of such models
are exactly known and are parameterized by a a set of quantum variables,
rapidities, $\left\{ k_{i}\right\} $. The equilibration or lack thereof
of some local observable $\Theta$ is captured by the time evolved
expectation value \cite{key-12,key-13,key-14,key-20,key-23,key-24,key-25,key-26,key-27,key-28,key-6,key-24-1,key-25-1,key-8,key-8-1}:
\begin{equation}
\begin{array}{l}
\left\langle \Theta\left(t\right)\right\rangle \equiv\left\langle \Phi\left(t=0\right)\right|e^{itH}\Theta e^{-itH}\left|\Phi\left(t=0\right)\right\rangle =\\
\qquad=\sum_{q}\sum_{k}\left\langle \Phi\left(t=0\right)\mid k\right\rangle \left\langle k\right|\Theta\left|q\right\rangle \times\\
\qquad\qquad\qquad\times\left\langle q\mid\Phi\left(t=0\right)\right\rangle e^{i\left(E_{k}-E_{q}\right)t}
\end{array}\label{eq:ExpectationValueOperator}
\end{equation}
Here $|k\rangle$ and $|q\rangle$ are complete sets of states and
$E_{k}$ and $E_{q}$ are their respective energies \cite{key-33-1}.

For translationally invariant systems, be they integrable or not,
it has been conjectured \cite{key-22,key-29,key-30,key-33} that at
large times, $t\rightarrow\infty$, the expectation value in Eq. (\ref{eq:ExpectationValueOperator})
equilibrates to a diagonal ensemble, where only entries with $\left|k\right\rangle =\left|q\right\rangle $
contribute. For non-integrable systems it has been further conjectured,
the ETH conjecture \cite{key-30,key-29}, that the expectation value
depends smoothly on the energy of the state $\{k\}$ and on no other
parameters, $\left\langle k\right|\Theta\left|k\right\rangle =F\left(E_{k}\right)$.
This conjecture leads directly to the fact that the long time limit
of $\left\langle \Theta\right\rangle $ may be computed in the microcanonical
ensemble. For integrable systems there is a different conjecture,
the GGE hypothesis \cite{key-20}, that the correlation function may
be computed using the GGE density matrix: $\rho_{GGE}=Z^{-1}\exp\left[-\sum_{m}\alpha_{m}I_{m}\right]$.
Here $I_{m}$ is the full set of commuting integrals of motion with
$|k\rangle$ being their common eigenstates, $I_{m}|k\rangle=\sum_{i}k_{i}^{m}|k\rangle$,
$Z=Tr\left[\exp\left(-\sum\alpha_{m}I_{m}\right)\right]$ is the partition
function and $\left\{ \alpha_{m}\right\} $ are Lagrange multipliers
fixed by the initial conditions $Tr\left[I_{m}\rho_{GGE}\right]=\left\langle I_{m}\right\rangle \left(t=0\right)$.
The expectation value of any local observable at large times is conjectured
to be given by $\left\langle \Theta\left(t\rightarrow\infty\right)\right\rangle =Tr\left[\rho_{GGE}\Theta\right]$.

\begin{figure}
\includegraphics[width=160bp]{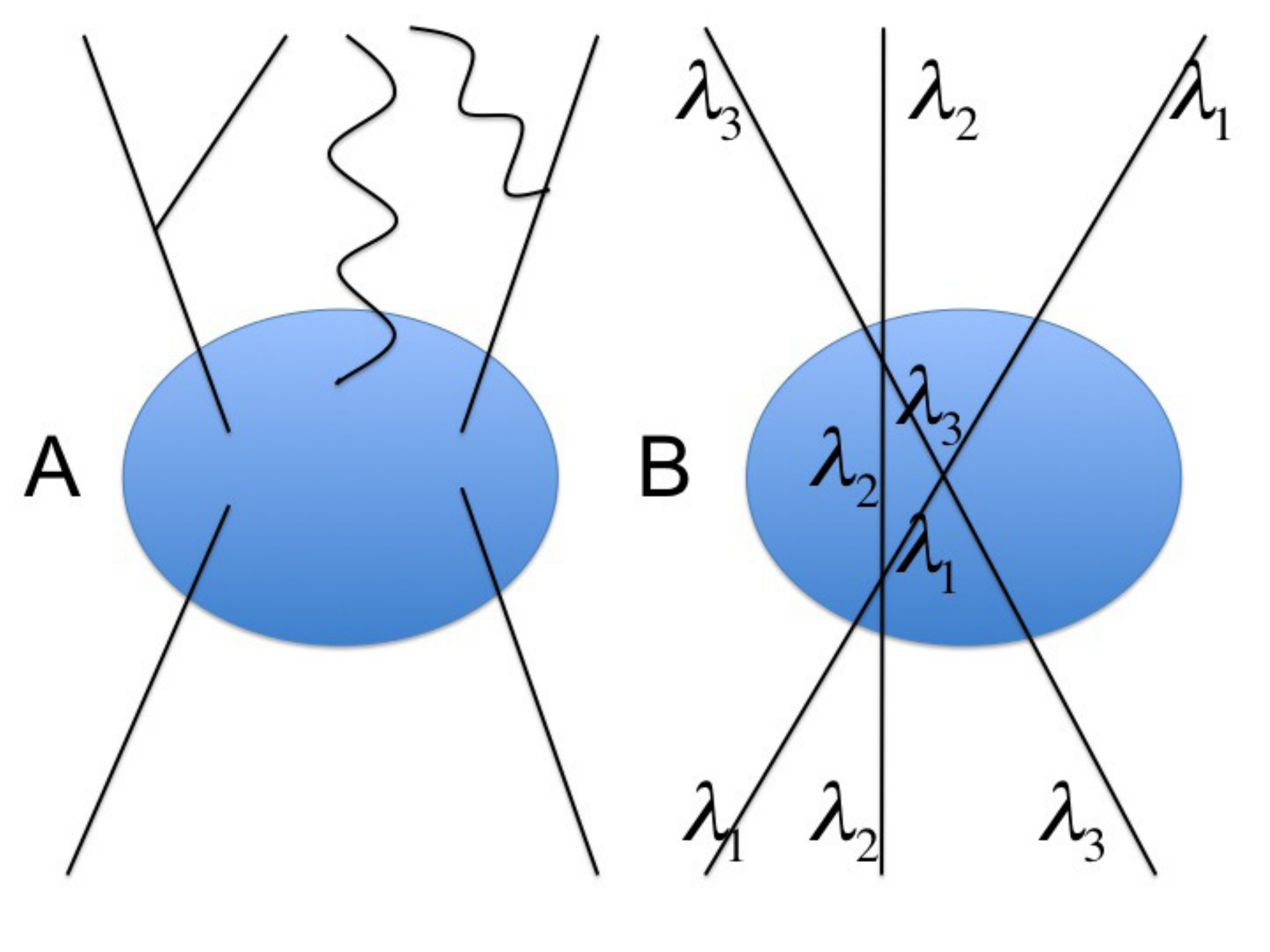}

\caption{\label{fig:scattering}Scattering. (A) For a non-integrable system
scattering leads to multi-particle production and decay so its impossible
to label the states by rapidities. (B) For an integrable system the
particle momenta do not change while scattering.}
\end{figure}

In this paper we examine these issues in the context of the Lieb-Liniger
model and provide explicit results for finite strong coupling as an
expansion in $1/c$, where $c$ is the lieb-liniger coupling constant,
see Eq. (\ref{eq:lieb-lin-hamiltonian}). We shall study the system
in the thermodynamic limit - where the system size $L\rightarrow\infty$,
the number of particles $N$ scales with the system size $N/L=const$,
and for times much less then the system size, $t<L/v_{typ}$ ($v_{typ}$
is a typical velocity). We show explicitly (1) that the system equilibrates
at long times, (2) that it is then described by a diagonal ensemble
and (3) that this ensemble is a \textit{ generalized} GGE, defined
as: 
\begin{equation}
\widehat{\rho}_{GGGE}=\widetilde{Z}^{-1}\exp\left[-\sum_{m_{1}m_{2}...}\alpha_{m_{1}m_{2}...}I_{m_{1}}I_{m_{2}}....\right]\label{eq:GGGE-Density-Matrix}
\end{equation}
Here $\widetilde{Z}=Tr\left[\exp\left(-\sum\alpha_{m_{1}m_{2}...}I_{m_{1}}I_{m_{2}}...\right)\right]$
with the Lagrange multipliers $\left\{ \alpha_{m_{1}m_{2}...}\right\} $
fixed by the initial conditions $Tr\left[I_{m_{1}}I_{m_{2}}...\rho_{GGGE}\right]=\left\langle I_{m_{1}}I_{m_{2}}...\right\rangle \left(t=0\right)$.
We note that products of the $I_{m}$ as used in Eq. (\ref{eq:GGGE-Density-Matrix})
are also conserved albeit nonlocal quantities. We show below that
when the initial state contains only short range correlations the
\textit{ generalized} GGE reduces to the usual GGE. The nonlocality
of the \textit{ generalized} GGE thus reflects the long range correlation
of the initial state when present. As we show below the \textit{generalized
GGE} appears when the long time limit of ensemble is diagonal and
the expectation value of a generic operator $\Theta$ may be Taylor
expanded in the rapidities: 
\begin{eqnarray*}
\left\langle k\right|\Theta\left|k\right\rangle =c_{0}+c_{1}\sum k_{i}+c_{1,1}\sum k_{i}k_{j}+c_{2}\sum k_{i}^{2}+..
\end{eqnarray*}
We establish the validity of this Taylor expansion for specific operators
below. We note that no such expansion may be made for non-integrable
models as there is no convenient set of rapidities to parameterize
them, see Fig. (\ref{fig:scattering}). For example for an electron
gas with coulomb interactions the eigenstates are complex combinations
of products of single particle states whose only constraint is to
have the same total momentum and energy. We finally note that for
strongly non-translationally invariant systems, such as those with
domain walls, the system never equilibrates in the thermodynamic limit
and in particular it does not attain the diagonal ensemble. Such a
state will be studied below as an example.

The simplest model that describes the dynamics of strongly correlated
1-d bosons, and which can also be realized in the lab is the Lieb-Liniger
Hamiltonian, 
\begin{equation}
H_{LL}=\intop_{-L/2}^{L/2}dx\left\{ \partial_{x}b^{\dagger}\left(x\right)\partial_{x}b\left(x\right)+c\left(b^{\dagger}\left(x\right)b\left(x\right)\right)^{2}\right\} \label{eq:lieb-lin-hamiltonian}
\end{equation}
Here $b^{\dagger}\left(x\right)$ is the bosonic creation operator
at the point $x$. The model is integrable \cite{key-3} and has infinitely
many conserved quantities. For the purposes of this work we shall
assume repulsive interactions, $c>0$. This parameter may be experimentally
tuned in real time via a Feshbach resonance. The equilibration of
this system starting from a highly excited state is one of the most
interesting properties of interacting bosonic many body systems and
it has been extensively studied \cite{key-12,key-13,key-14,key-20,key-23,key-24,key-25,key-26,key-27,key-28,key-24-1,key-25-1,key-8,key-8-1}.

There are varied techniques to study analytically the quench dynamics
and equilibration of such models. Most of these are based on the fact
that it is possible to find exact eigenstates of the many body Hamiltonian,
decompose initial states in terms of these eigenstates and then time
evolve, see Eq. (\ref{eq:ExpectationValueOperator}). As such, from
the theory side, the study of the evolution and the equilibration,
of correlation functions for an initial state may be decomposed into
four steps. (1) The computation of the eigenstates $\left|k\right\rangle $
of an exactly integrable system; which may be done using co-ordinate
Bethe Ansatz techniques \cite{key-16,key-31}. (2) The computation
of various overlaps $\left\langle \Phi\left(t=0\right)\mid k\right\rangle $;
the Yudson representation, which we shall extend to finite size systems,
is an efficient technique for accomplishing almost that. (3) The computation
of various matrix elements for local operators $\left\langle k\right|\Theta\left|q\right\rangle $,
(4) Summation (or integration, in the thermodynamic limit) over various
intermediate states which for the strong coupling limit, as we shall
show, may be converted into a computation of an appropriate correlation
function with respect to the initial state.

In this paper we describe how to accomplish these four steps for generic
initial states, be they translationally invariant or not, with short
or long range correlations. We present generic formulas for the correlation
functions for these states at arbitrary times in terms of correlation
functions of the initial state. We show that in the long time thermodynamic
limit the correlation functions equilibrate and for translationally
invariant initial states they equilibrate to a diagonal ensemble,
that is $\left|k\right\rangle =\left|q\right\rangle $ in Eq. (\ref{eq:ExpectationValueOperator})
above. We shall also show the validity of the GGGE hypothesis. As
far as the authors are aware this is the first analytic proof of equilibration,
diagonal ensemble or GGGE for the interacting case see however \cite{key-40}.

\textit{Yudson decomposition} provides an efficient way to compute
overlaps. We extend previous studies \cite{key-2,key-4,key-5} to
finite size systems, an extension that is necessary to reach the thermodynamic
limit. The Yudson decomposition for a $N$-particles state on a ring
of length $L$ may be described as follows: supposing that our initial
state may be written as an integral over a wave function localized
in the first quadrant of our co-ordinate space $x_{1}<x_{2}<....<x_{N}$
(from Bose symmetry it follows that every function may be written
like that): 
\[
\begin{array}{l}
\left|\Phi_{N}\left(t=0\right)\right\rangle =\intop_{-L/2}^{L/2}dx_{N}\intop_{-L/2}^{x_{N}}dx_{N-1}....\intop_{-L/2}^{x_{2}}dx_{1}\times\\
\qquad\Phi\left(x_{1},x_{2}...x_{N}\right)b^{\dagger}\left(x_{N}\right)....b^{\dagger}\left(x_{1}\right)\left|0\right\rangle ,
\end{array}
\]
then we claim this can be written as a sum of Bethe ansatz eigenstates
of the form: 
\begin{equation}
\begin{array}{l}
\left|\Phi_{N}\left(t=0\right)\right\rangle =\sum_{n_{1}=-\infty}^{\infty}...\sum_{n_{N}=-\infty}^{\infty}\mathcal{N}\left(k_{n_{1}}....k_{n_{N}}\right)^{-1}\times\\
\qquad\times\left|k_{n_{1}}....k_{n_{N}}\right\rangle \left(k_{n_{1}}...k_{n_{N}}\right|\left|\Phi_{N}\left(t=0\right)\right\rangle .
\end{array}\label{eq:YudsonDecomposition}
\end{equation}
Here $\left|k_{n_{1}}...k_{n_{N}}\right\rangle $ is a Bethe ansatz
eigenstate: 
\begin{equation}
\begin{array}{l}
\intop_{-L/2}^{L/2}dy_{N}...\intop_{-L/2}^{L/2}dy_{1}\times\\
\qquad\times\prod_{i<j}Z_{y_{i}-y_{j}}\left(k_{i}-k_{j}\right)\cdot\prod e^{ik_{i}y_{i}}\cdot\prod b^{\dagger}\left(y_{i}\right)\left|0\right\rangle 
\end{array}\label{eq:Yudson_eigenstate}
\end{equation}
where the scattering factor $Z_{Y}\left(K\right)\equiv\frac{K+ic\left(1-2\theta\left(Y\right)\right)}{K+ic}$
incorporates the $S$-matrix, $S_{ij}=\frac{k_{i}-k_{j}+ic}{k_{i}-k_{j}+ic}$.
With periodic boundary conditions the momenta $\left\{ k_{n_{i}}\right\} $
must satisfy the Bethe ansatz equations: $k_{n_{i}}=\frac{2\pi}{L}n_{i}-2\sum_{l=1}^{N}\arctan\left(\frac{2\pi}{L}\cdot\frac{k_{i}-k_{l}}{c}\right),\forall i,$
with $n_{i}$ (half) integers. We denote: $\left|k_{n_{1}}...k_{n_{N}}\right)=\int_{x}\prod e^{ik_{n_{i}}x_{i}}\prod b^{\dagger}\left(y_{i}\right)\left|0\right\rangle $.
The inner product is taken over one quadrant only $\left(\right|\left|\right\rangle =\intop_{-L/2}^{L/2}dx_{N}\intop_{-L/2}^{x_{N}}dx_{N-1}...\intop_{-L/2}^{x_{2}}dx_{1}$.
The normalization of the wave function is $\mathcal{N}\left(k_{n_{1}}....k_{n_{N}}\right)=\det\left(M_{jk}\right)$
with $M_{jk}=\delta_{jk}\left(L+\sum_{l=1}^{N}\frac{2c}{c^{2}+\left(k_{j}-k_{l}\right)^{2}}\right)-\frac{2c}{c^{2}+\left(k_{j}-k_{k}\right)^{2}}$,
see \cite{key-1}. In the limit $L\rightarrow\infty$ for a finite
number of particles this simplifies to $M_{jk}=L\delta_{jk}$ and
$\mathcal{N}\left(k_{n_{1}}....k_{n_{N}}\right)=L^{N}$. The density
of states becomes $\left(\frac{L}{2\pi}\right)^{N}$. The proof of
the Yudson resolution of the identity $\mathbb{I}_{N}=\sum_{n_{1},n_{2},...n_{N}}\frac{1}{\mathcal{N}\left(k_{n_{1}}....k_{n_{N}}\right)}\left|k_{n_{1}},...k_{n_{N}}\right\rangle \left(k_{n_{1}},....k_{n_{N}}\right|$
follows from the standard resolution: $\mathbb{I}_{N}=\sum_{n_{1}<n_{2}<...n_{N}}\frac{1}{\mathcal{N}\left(k_{n_{1}}....k_{n_{N}}\right)}\left|k_{n_{1}},...k_{n_{N}}\right\rangle \left\langle k_{n_{1}},....k_{n_{N}}\right|$
and the identity: $\left\langle k_{1},...k_{N}\right|=\sum_{P\subset S_{N}}\left(k_{P1},...k_{PN}\right|\prod_{i,j\in P}S^{\star}\left(k_{i},k_{j}\right)$
valid in the first quadrant \cite{key-38}. It is also useful to express
the Yudson resolution in terms of Algebraic Bethe Ansatz (ABA) states
$B\left(k_{1}\right)...B\left(k_{N}\right)\left|0\right\rangle $
(they are proportional to $\left|k_{n_{1}},...k_{n_{N}}\right\rangle $
see \cite{key-32-1}) with the resolution taking the form: 
\begin{eqnarray*}
 & \mathbb{I}_{N}=\sum_{n_{1},n_{2},...n_{N}}\frac{1}{\mathcal{N}\left(k_{n_{1}}....k_{n_{N}}\right)}\times\quad\\
 & \frac{\prod_{i<j}\left(k_{n_{j}}-k_{n_{i}}\right)}{\left(-i\sqrt{c}\right)^{N}\prod_{i<j}\left(k_{n_{j}}-k_{n_{i}}-ic\right)}B\left(k_{n_{1}}\right)..B\left(k_{n_{N}}\right)\left|0\right\rangle \left(k_{n_{1}},...k_{n_{N}}\right|
\end{eqnarray*}

\textit{Greens Functions.} In order to calculate the expectation values
of $\left\langle \Theta\left(t\right)\right\rangle $, see Eq. (\ref{eq:ExpectationValueOperator}),
we wish to calculate the value of operator Greens functions in the
basis states: 
\begin{equation}
\begin{array}{l}
G\left(\Theta,t;x_{1},x_{2},....x_{N};y_{1},...y_{N}\right)=\\
\left\langle 0\right|b\left(y_{1}\right)b\left(y_{2}\right)...b\left(y_{N}\right)\Theta\left(t\right)b^{\dagger}\left(x_{1}\right)...b^{\dagger}\left(x_{N}\right)\left|0\right\rangle 
\end{array}\label{eq:OperatorGreensFunctions}
\end{equation}
This allows a basis for calculating the expectation $\left\langle \Theta\left(t\right)\right\rangle $
with any initial and final states $\Psi,\,\Phi$ since: 
\begin{equation}
\begin{array}{l}
\left\langle \Psi\right|\Theta\left(t\right)\left|\Psi\right\rangle =\int...\int dx_{1}....dx_{N}dy_{1}...dy_{N}\times\\
\quad\times G\left(\Theta,t;x_{1}...x_{N};y_{1}....y_{N}\right)\Psi\left(x_{1},....x_{N}\right)\Phi^{*}\left(y_{1}...y_{N}\right)
\end{array}\label{eq:greens_basis}
\end{equation}
Using the Yudson representation we may rewrite the Green's functions
in the form: 
\begin{equation}
\begin{array}{l}
G^{LL}\left(\Theta,t;x_{1},x_{2},....x_{N};y_{1},...y_{N}\right)=\\
\sum_{n_{1},n_{2},...n_{N}}\frac{1}{\mathcal{N}\left(k_{n_{1}}....k_{n_{N}}\right)}\times\frac{\prod_{i<j}\left(k_{n_{j}}-k_{n_{i}}\right)}{\left(i\sqrt{c}\right)^{N}\prod_{i<j}\left(k_{n_{j}}-k_{n_{i}}+ic\right)}\times\\
\times\left\langle 0\right|b\left(y_{1}\right)b\left(y_{2}\right)...b\left(y_{N}\right)\left|k_{n_{1}}...k_{n_{N}}\right)\times\\
\times\left\langle B\left(k_{n_{1}}\right)....B\left(k_{n_{N}}\right)\right|\Theta\left|B\left(q_{n_{1}}\right),...B\left(q_{n_{N}}\right)\right\rangle \times\\
\times\sum_{n_{1},n_{2},...n_{N}}\frac{1}{\mathcal{N}\left(q_{n_{1}}....q_{n_{N}}\right)}\times\frac{\prod_{i<j}\left(q_{n_{j}}-q_{n_{i}}\right)}{\left(-i\sqrt{c}\right)^{N}\prod_{i<j}\left(q_{n_{j}}-q_{n_{i}}-ic\right)}\\
\left(q_{n_{1}},....q_{n_{N}}\right|b^{\dagger}\left(x_{1}\right)...b^{\dagger}\left(x_{N}\right)\left|0\right\rangle \prod_{i}e^{i\left(k_{n_{i}}-q_{n_{i}}\right)t}
\end{array}\label{YudsonGreensFunctions-Lieb-Lininger-1}
\end{equation}

We note that $\left\langle 0\right|b\left(y_{1}\right)b\left(y_{2}\right)...b\left(y_{N}\right)\left|k_{n_{1}}...k_{n_{N}}\right)=\prod\exp\left(ik_{n_{i}}y_{i}\right)$
with $x_{1}...x_{N}$ and $y_{1}...y_{N}$ in the first quadrant.
We will consider the operator $\Theta=\exp\left(\alpha Q_{xy}\right)\equiv\exp\left(\alpha\int_{x}^{y}b^{\dagger}\left(z\right)b\left(z\right)dz\right)$,
from which all local density $\rho(x)=b^{\dagger}(x)b(x)$ correlation
functions can be obtained, e.g. $\rho\left(x\right)\rho\left(y\right)=-\frac{1}{2}\frac{\partial^{2}}{\partial x\partial y}\frac{\partial^{2}}{\partial\alpha^{2}}\exp\alpha Q_{xy}\left(\alpha=0\right)$.
As $\left\langle B\left(k_{n_{1}}\right)....B\left(k_{n_{N}}\right)\right|\Theta\left|B\left(q_{n_{1}}\right),...B\left(q_{n_{N}}\right)\right\rangle $
may be efficiently calculated see \cite{key-1} it is possible to
know the exact correlators using only a finite Taylor expansion with
respect to $\alpha$. Expanding the expectation value we find, after
a considerable amount of algebra, in the thermodynamic limit the generating
function is given by: 
\begin{equation}
\begin{array}{l}
\left\langle \exp\left(\alpha Q_{xy}\left(t\right)\right)\right\rangle =1+\int dXdYF_{\alpha,xy}\left(X,Y,t\right)\times\\
\qquad\times\left\langle b^{\dagger}\left(Y\right)\exp\left[i\int_{X}^{Y}dz\pi b^{\dagger}\left(z\right)b\left(z\right)\right]b\left(X\right)\right\rangle +\\
+\int dX_{1}dX_{2}dY_{1}dY_{2}\times F_{\alpha,xy}\left(X_{1},Y_{1},t\right)F_{\alpha,xy}\left(X_{2},Y_{2},t\right)\times\\
\times\left\langle sgn\left(Y_{2}-Y_{1}\right)b^{\dagger}\left(Y_{1}\right)b^{\dagger}\left(Y_{2}\right)e^{i\int_{X_{1}}^{Y_{1}}dz\pi b^{\dagger}\left(z\right)b\left(z\right)}\times\right.\\
\times\left.sgn\left(X_{2}-X_{1}\right)e^{i\int_{X_{2}}^{Y_{2}}dz\pi b^{\dagger}\left(z\right)b\left(z\right)}b\left(X_{1}\right)b\left(X_{2}\right)\right\rangle -\\
-\frac{i\alpha}{\pi^{2}c}\int dX_{1}dY_{1}dX_{2}dY_{2}\left\{ F_{\alpha,x}\left(X_{1},Y_{1},t\right)G_{\alpha,x}\left(X_{2},Y_{2},t\right)-\right.\\
\left.-F_{\alpha,y}\left(X_{1},Y_{1},t\right)G_{\alpha,y}\left(X_{2},Y_{2},t\right)\right\} \left\langle sgn\left(y_{2}-y_{1}\right)\right.\\
\cdot sgn\left(x_{2}-x_{1}\right)\cdot b^{\dagger}\left(y_{1}\right)b^{\dagger}\left(y_{2}\right)e^{i\int_{X_{1}}^{Y_{1}}dz\pi b^{\dagger}\left(z\right)b\left(z\right)}\cdot\\
\left.\cdot e^{i\int_{X_{2}}^{Y_{2}}dz\pi b^{\dagger}\left(z\right)b\left(z\right)}b\left(X_{1}\right)b\left(X_{2}\right)\right\rangle -\\
-\frac{i\alpha}{2\pi^{2}c}\int dXdYG_{\alpha,X}\left(X,Y,t\right)\left\langle b^{\dagger}\left(Y\right)e^{i\int_{X}^{Y}dz\pi b^{\dagger}\left(z\right)b\left(z\right)}\cdot\right.\\
\left.\cdot\int_{-\infty}^{\infty}dvsgn\left(x-v\right)\rho\left(v\right)sgn\left(v-Y\right)sgn\left(v-X\right)\right\rangle +\\
+\frac{i\alpha}{2\pi^{2}c}\int dXdYG_{\alpha,y}\left(X,Y,t\right)\left\langle b^{\dagger}\left(Y\right)e^{i\int_{X}^{Y}dz\pi b^{\dagger}\left(z\right)b\left(z\right)}\cdot\right.\\
\left.\cdot\int_{-\infty}^{\infty}dvsgn\left(y-v\right)\rho\left(v\right)sgn\left(v-Y\right)sgn\left(v-X\right)\right\rangle +....
\end{array}\label{eq:Taylor_expand_Lieb_liniger}
\end{equation}

Here we have introduced $F_{\alpha,xy}\left(X,Y,t\right)\equiv i\frac{e^{\alpha}-1}{2\pi}\frac{\exp\left(-\frac{i}{4t}\left(Y^{2}-X^{2}\right)\right)}{Y-X}\times\left[\exp\left(\frac{i\left(Y-X\right)\cdot x}{2t}\right)-\exp\left(\frac{i\left(Y-X\right)\cdot y}{2t}\right)\right]$,
$G_{\alpha,x}\left(X,Y,t\right)\equiv\exp\left(i\frac{\left(Y-X\right)x}{2t}\right)\frac{\exp\left(-i\left(Y^{2}-X^{2}\right)/2t\right)}{t}$,
$G_{\alpha,y}\left(X,Y,t\right)\equiv\exp\left(i\frac{\left(Y-X\right)y}{2t}\right)\frac{\exp\left(-i\left(Y^{2}-X^{2}\right)/2t\right)}{t}$,
$F_{\alpha,x}\left(X,Y,t\right)\equiv\frac{\exp\left(-\frac{i}{4t}\left(Y^{2}-X^{2}\right)\right)}{Y-X}\exp\left(\frac{i\left(Y-X\right)\cdot x}{2t}\right)$,
and $F_{\alpha,y}\left(X,Y,t\right)\equiv\frac{\exp\left(-\frac{i}{4t}\left(Y^{2}-X^{2}\right)\right)}{Y-X}\exp\left(\frac{i\left(Y-X\right)\cdot y}{2t}\right)$.
This expression yields the density and density density correlation
functions to leading order for arbitrary times and positions, valid
for an arbitrary initial state.

\textit{Long time thermodynamic limit of $G\left(\exp\left(\alpha Q_{x,y}\right),t\right)$.}
We would like to show that for any initial state, for which various
field correlation functions have convergent smooth Fourier transforms,
the expectation value of $\left\langle \exp\alpha Q_{xy}\right\rangle $
converges to a constant value in the thermodynamic limit. For this
we will consider Eq. (\ref{eq:Taylor_expand_Lieb_liniger}), and note
that the Fourier transform of all the functions $F/G_{\alpha,xy/x/y}\left(X,Y,t\right)$
with respect to $X,Y$ are proportional to $e^{i\left(q^{2}-k^{2}\right)t}$
\cite{key-34-1}. These terms are multiplied by the expectation values
of the Fourier transforms of some correlations functions, e.g. the
Fourier transform of the terms in the correlation functions $\left\langle \right\rangle $
in Eq. (\ref{eq:Taylor_expand_Lieb_liniger}). Call these Fourier
transforms $\mathrm{O}_{n}\left(\left\{ k_{i}\right\} ,\left\{ q_{i}\right\} \right)$.
If $\mathrm{O}_{n}\left(\left\{ k_{i}\right\} ,\left\{ q_{i}\right\} \right)$
is smooth then from the method of stationary phase we know that any
integral containing this term is dominated by the point $k_{i}=q_{i}=0$
and is proportional $\propto\frac{1}{t^{n}}$ \cite{key-19}, so it
disappears in the long time limit. Therefore to get a non-zero result
at long times we need to assume that the Fourier transform $\mathrm{O}_{n}\left(\left\{ k_{i}\right\} ,\left\{ q_{i}\right\} \right)$
has singularities proportional to delta functions with the support
of these singularities being the set where $\sum k_{i}^{2}-\sum q_{i}^{2}=const$.
However when this condition is satisfied the expression for $\left\langle \exp\left(\alpha Q_{xy}\left(t\right)\right)\right\rangle $
explicitly has no time dependance and therefore equilibrates. Singularities
in Fourier space are determined by the initial state and correspond
to some order in it. For most initial states. e.g. translationally
invariant states, lattices, superconducting order etc., these ordering
are localized to q-vectors that lie in hyperplanes - not curved manifolds.
For hyperplanes the condition $\sum k_{i}^{2}-\sum q_{i}^{2}=const$
simplifies to $k_{i}=\pm q_{j}$. Furthermore for states that have
no ordering at non-zero total momentum, such as translationally invariant
states, this constraint simplifies to $k_{i}=q_{j}$. In this case
the long time thermodynamic limit is given by the diagonal ensemble.
Indeed the condition $k_{i}=q_{j}$ implies that $\left\{ k_{i}\right\} =\left\{ q_{j}\right\} $
or the rapidities of the states used in the $ $Yudson decomposition
in Eq. (\ref{YudsonGreensFunctions-Lieb-Lininger-1}) are the same.
Our derivation applies also to states with crystal order, these do
have ordering at non-zero momentum but the extra singularities in
the Green's functions are hyperplanes that do not pass through the
origin. This result also applies to states with a finite number of
defects as the Fourier transform of a defect is a smooth function
so it contribution is $\propto\frac{1}{t^{n}}$. We note that the
leading term correction to the steady state is given when two of the
rapidities $k_{i}\neq q_{j}$ and the rest are the same. In this case
by the method of stationary phase we see that the corrections to the
stationary results are given by $\sim\frac{Const}{t}+\frac{Const}{t^{2}}+...$
We note that this time dependance scaling derivation works only for
systems with some disorder, non-zero $\mathrm{O}_{n}\left(\left\{ k_{i}\right\} ,\left\{ q_{i}\right\} \right)$
near $k_{i}=q_{j}=0$. In particular for initial states with perfect
crystal oder the decay can be exponential see the discussion above
Eq. (\ref{eq:Density}). A nearly identical proof of equilibration
applies to : $\lim_{t\rightarrow\infty}\left\langle e^{\alpha_{1}Q_{x_{1}y_{1}}\left(t+t_{1}\right)}....e^{\alpha_{n}Q_{x_{n}y_{n}}\left(t+t_{n}\right)}\right\rangle ,$
\textit{i.e. }any correlation function of density operators equilibrates
\cite{key-18}.

\textit{The GGGE ensemble.} We now show that the equilibrated system
is described by a generalized GGE. This follows from from the fact
that we can Taylor expand observables, $\left\langle \left\{ k_{i}\right\} \right|\Theta\left|\left\{ k_{i}\right\} \right\rangle =c_{0}+c_{1}\sum k_{i}+c_{1,1}\sum k_{i}k_{j}+c_{2}\sum k_{i}^{2}+..$,
and that the ensemble turn out to be diagonal in the thermodynamic
limit. The Taylor expansion follows by inspection to any order in
$1/c$ from its expression see \cite{key-1}. To derive the GGGE consider
an arbitrary diagonal density matrix $\rho_{D}=\sum p_{\left\{ k\right\} }\left|\left\{ k_{i}\right\} \right\rangle \left\langle \left\{ k_{i}\right\} \right|$,
(in our case, $\rho_{D}=\sum_{\lambda}tr\left[\rho\left(t=0\right)\left|k\right\rangle \left\langle k\right|\right]\left|\left\{ k_{i}\right\} \right\rangle \left\langle \left\{ k_{i}\right\} \right|$),
then introducing $\left\langle I_{1}\right\rangle =\sum p_{\left\{ k\right\} }\sum k_{i}$,
$\left\langle I_{1}^{2}\right\rangle =\sum p_{\left\{ k\right\} }\sum k_{i}k_{j}$,
$\left\langle I_{2}\right\rangle =\sum p_{\left\{ k\right\} }\sum k_{i}^{2}$,
we see that $\left\langle \Theta\right\rangle =Tr\rho_{D}\Theta=c_{0}+c_{1}\left\langle I_{1}\right\rangle +c_{1,1}\left\langle I_{1}^{2}\right\rangle +c_{2}\left\langle I_{2}\right\rangle +..$.
From this expression we see that the expectation value of any operator
with a Taylor expansion is determined by the values $\left\langle I_{1}\right\rangle $,
$\left\langle I_{1}^{2}\right\rangle $, $\left\langle I_{2}\right\rangle $
and so forth. However the GGGE ensemble has exactly the same expectation
value for these operators as the diagonal ensemble, see the discussion
following Eq. (\ref{eq:GGGE-Density-Matrix}) so the expectation value
of any Taylor expandable operator - which includes all correlation
functions of densities \cite{key-18} - may be computed in GGGE. It
is clear that the GGGE density operator takes the form $\hat{\rho}=\int dkf(k)|k\rangle\langle k|$
while the GGE density matrix corresponds to a single eigenvalue of
the Lieb-Liniger gas \cite{key-37}. Thus for the GGE $\left\langle I_{m_{1}}I_{m_{2}}....\right\rangle =\left\langle I_{m_{1}}\right\rangle \left\langle I_{m_{2}}\right\rangle ...$
and in particular 
\begin{equation}
\begin{array}{l}
\left\langle \Theta\left(t\rightarrow\infty\right)\right\rangle -tr\left[\Theta\rho_{GGE}\right]=\\
=c_{1,1}\left(\left\langle I_{1}^{2}\right\rangle -\left\langle I_{1}\right\rangle ^{2}\right)+c_{1,2}\left(\left\langle I_{1}I_{2}\right\rangle -\left\langle I_{1}\right\rangle \left\langle I_{2}\right\rangle \right)+....
\end{array}\label{eq:Difference_GGE}
\end{equation}
implying that long range correlations are not captured by GGE. An
example of an initial state that does not have $\left\langle I_{2}^{2}\right\rangle \neq\left\langle I_{2}\right\rangle ^{2}$
is $\rho_{T}=\frac{1}{T}\int_{0}^{T}\frac{e^{-H_{LL}/t}}{Z_{t}}$
and it is also possible to construct examples involving pure states\cite{key-18}.
However for many initial states $\lim_{\left|x-y\right|\rightarrow\infty}\left\langle J_{i}\left(y\right)J_{j}\left(x\right)\right\rangle =\left\langle J_{i}\left(x\right)\right\rangle \left\langle J_{j}\left(y\right)\right\rangle $,
etc. in which case the GGGE reduces to the GGE. Here $J_{i}\left(x\right)$
are the local densities corresponding to $I_{i}$, $I_{i}=\int J_{i}\left(x\right)$.
As such we have shown that for translationally invariant systems correlations
of densities equilibrate to the diagonal ensemble so they automatically
equilibrate to the GGGE.

\begin{figure}
\begin{centering}
\includegraphics[width=1\columnwidth]{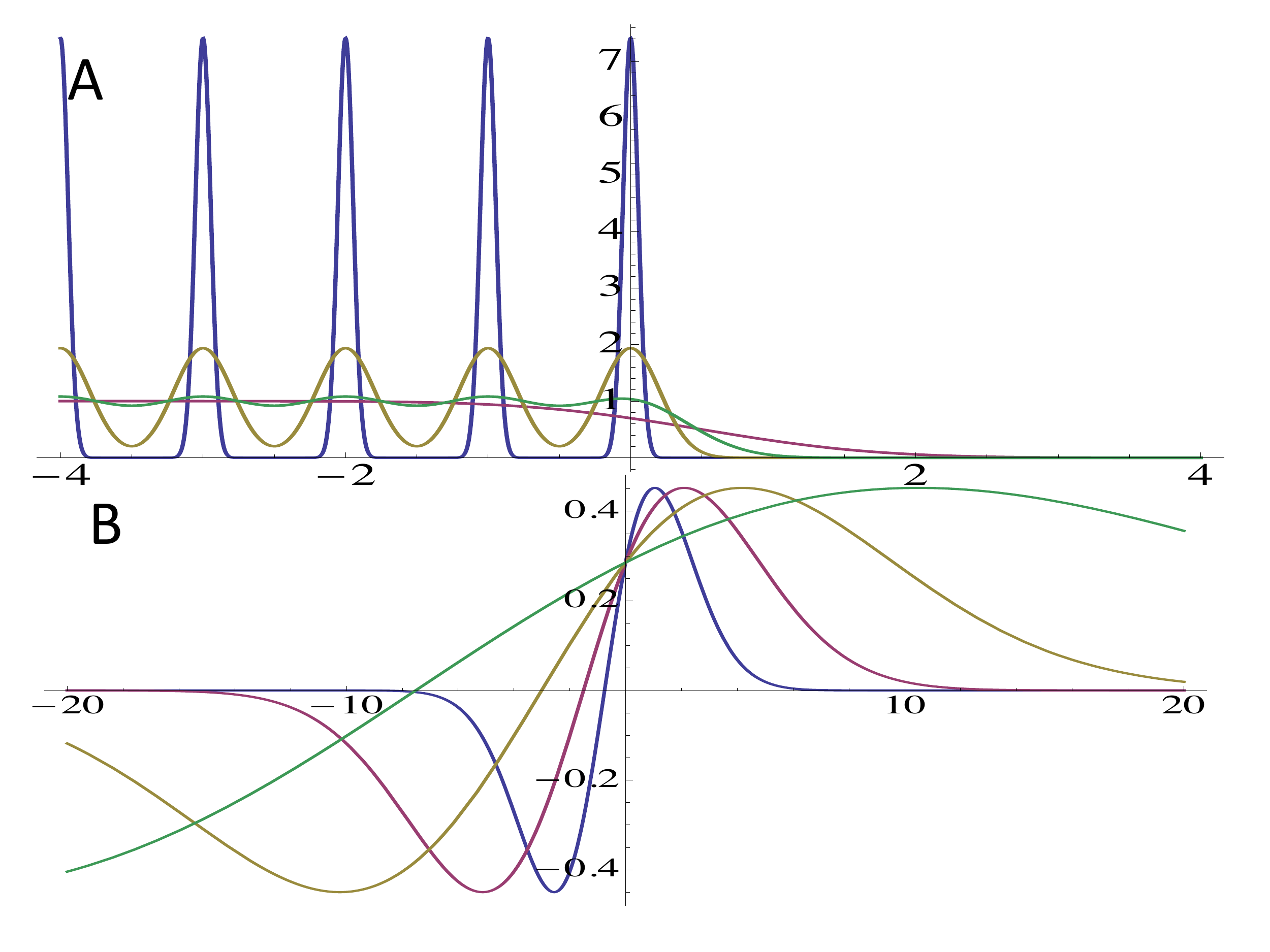} 
\par\end{centering}

\caption{\label{fig:Initial_States}For initial state a Mott insulator on the
half line: correlation functions: (A) For a the Tonks Gas ($c\rightarrow\infty$)
exact correlations near the edge of the insulator (B) leading order
$1/c$ corrections for the edge of the insulator, the entire graph
(B) should be multiplied by $\frac{16}{\pi c}$. The initial state
is chosen to have parameters $l=1$ $\sigma=0.01$ and the times are
$0.001$, $0.01$, $0.02$ and $0.05$ for the Tonks Girardeau gas
and $0.1$, $0.2$, $0.4$ and $1$ for the $1/c$ corrections.}
\end{figure}

\textit{Examples of initial states.} We would like to apply our formalism
to some interesting initial states. We shall study a quench from from
an initial state whose average density is of the form of a domain
wall, see Fig. (\ref{fig:Initial_States}), and study the density
evolution as a function of time $\rho\left(x,t\right)$. Choosing
an initial state that is a Mott insulator on the half line: $\left|\Psi\left(t=0\right)\right\rangle =\prod_{j=0}^{\infty}\int_{-\infty}^{\infty}\varphi\left(x+jl\right)b^{\dagger}\left(x\right)\left|0\right\rangle $,
with $\varphi\left(x\right)=\frac{e^{-x^{2}/\sigma}}{\left(\pi\sigma/2\right)^{1/4}}$,
will allow us to demonstrate both equilibration by considering $x\to-\infty$
and nonequilibrium ballistic transport physics for $\left|x\right|\sim t$.
For $x\to-\infty$ in the Tonks Girardeau regime it is possible to
compute the density exactly. It is given by $\rho\left(x,t\right)=\frac{1}{l}\left(1+\sum_{s=1}^{\infty}e^{-\pi^{2}\left(8t^{2}/\sigma+\sigma/2\right)s^{2}/l^{2}}\cos\left(2\pi sx/l\right)\right)$.
At large times this corresponds to a constant density plus exponentially
decaying small oscillations, see Fig. (\ref{fig:Initial_States}).
The $1/c$ correction corresponds to exponentially decaying small
oscillations. For $x\sim t$, ignoring exponentially decaying small
oscillating terms, we have that the density is given by: 
\begin{equation}
\begin{array}{l}
\rho\left(x,t\right)=\frac{1}{l}\left(\frac{1}{2}Erfc\left(\frac{x}{\sqrt{A}}\right)+\right.\\
+\frac{16}{\pi cl}e^{-x^{2}/A}\left(\frac{1}{2}\sqrt{\pi}\frac{x}{\sqrt{A}}Erfc\left(\frac{x}{\sqrt{A}}\right)-\frac{1}{2}e^{-x^{2}/A}\right)\\
\left.+\frac{16}{\pi cl}\frac{\pi}{2}\left(1-\frac{1}{2}Erfc\left(\frac{x}{\sqrt{A}}\right)\right)Erfc\left(\frac{x}{\sqrt{A}}\right)\right)
\end{array}\label{eq:Density}
\end{equation}
with $A\equiv8t^{2}/\sigma+\sigma/2$. We notice that for $x\ll-\sqrt{8t^{2}/\sigma}$
the density becomes $1/l$ or its equilibrium value while for $x\gg\sqrt{8t^{2}/\sigma}$
the density becomes zero corresponding to no particles having reached
our observation point and for $\left|x\right|\ll t$ the density becomes
$\frac{1}{l}\left(\frac{1}{2}+\frac{4\pi}{cl}\right)$. From this
we see ballistic transport with signal velocity $\sim\frac{1}{\sqrt{\sigma}}$.
We note that technically this system never equilibrates, for large
enough $x$ there is always time dependance, so in particular it does
not attain the GGE. A translationally invariant initial state, e.
g. $\left|\Psi\left(t=0\right)\right\rangle =\prod_{j=-\infty}^{\infty}\int_{-\infty}^{\infty}\varphi\left(x+jl\right)b^{\dagger}\left(x\right)\left|0\right\rangle $
equilibriates to a GGE, since $\left\langle I_{i}I_{j}...\right\rangle =\left\langle I_{i}\right\rangle \left\langle I_{j}\right\rangle ...$

\textit{Conclusions.} By studying the relation between differently
ordered Bethe eigenstates and decomposing a Bethe eigenstate appropriately
in terms of plane waves we have introduced a new type of Yudson representation,
valid for finite sized systems. We have shown how it can be used to
study the quench dynamics, in particular equilibration, of macroscopic
systems in the thermodynamic limit. We have introduced some techniques
for writing time dependent observables in terms of Green's functions
and initial correlation functions and we have demonstrated how to
take the long time thermodynamic limit of these functions. We have
used this technique to study the dynamics of the Lieb Liniger gas
and showed that for translationally invariant systems the gas equilibrates
to GGGE. We are currently applying our approach to the XXZ model,
the Gaudin-Yang model and to the Hubbard models.

\textbf{Acknowledgments}: This research was supported by NSF grant
DMR 1006684 and Rutgers CMT fellowship. We would like to thank D.
Iyer for useful discussions.

\part*{Supplementary online information}

We now wish to extend the Yudson decomposition for repulsive bosons
to systems of a finite size. That is we would like to prove that for
a finite sized system with periodic boundary conditions there is an
alternative resolution of unity:

\begin{equation}
\mathbb{I}_{N}=\sum_{n_{1},n_{2},...n_{N}}\frac{1}{\mathcal{N}\left(k_{n_{1}}....k_{n_{N}}\right)}\left|k_{n_{1}},...k_{n_{N}}\right\rangle \left(k_{n_{1}},....k_{n_{N}}\right|\mid\label{eq:resolutionofidentityfinitesizefinal-1-1}
\end{equation}
Here the normalization factor $\mathcal{N}\left(k_{n_{1}}....k_{n_{N}}\right)=\left\langle k_{n_{1}}...k_{n_{N}}\mid k_{n_{1}}...k_{n_{N}}\right\rangle $,
the rapidities $k_{i}$ are assumed to satisfy the Bethe ansatz equation,
the ket $\left|k_{n_{1}}...k_{n_{N}}\right\rangle $ is an exact eigenstate
of the Lieb Liniger hamiltonian, the bra $\left(k_{n_{1}}....k_{n_{N}}\right|$
is a plane wave state with $\left(k_{n_{1}}...k_{n_{N}}\right|\left(x_{1},..x_{N}\right)=\prod e^{-ik_{i}x_{i}}$
and the inner product is over the first quadrant e.g. $\left\langle \Phi\mid\mid\Psi\right\rangle =\int_{-L/2}^{L/2}\int_{-L/2}^{x_{N}}...\int_{-L/2}^{x_{2}}\Phi^{\star}\left(x_{1}...x_{N}\right)\Psi\left(x_{1}...x_{N}\right)$.
The representation will be based on the following resolution of identity:
\begin{equation}
\mathbb{I}_{N}=\sum_{n_{1}<n_{2}<...n_{N}}\frac{1}{\mathcal{N}\left(k_{n_{1}}....k_{n_{N}}\right)}\left|k_{n_{1}},...k_{n_{N}}\right\rangle \left\langle k_{n_{1}},....k_{n_{N}}\right|,\label{eq:finitesizeresolutionofidentity-1}
\end{equation}
which comes from the completeness of the bethe ansatz eigenstates.
To show that the two resolutions Eqs. (\ref{eq:resolutionofidentityfinitesizefinal-1-1})
and (\ref{eq:finitesizeresolutionofidentity-1}) are equal we will
need to recall some facts about the eigenstates given in Eq. (\ref{eq:Yudson_eigenstate}).
First in the quadrant $x_{1}<x_{2}<....<x_{N}$ they have a simple
form: 
\begin{equation}
\left|k_{n_{1}}....k_{n_{N}}\right\rangle =\sum_{P\in S_{N}}\prod_{\left(i,j\right)\in P}S\left(k_{i},k_{j}\right)\left|k_{P_{1}}...k_{P_{N}}\right)\label{eq:Eigenstates_quadrant}
\end{equation}
The second observation we need to make is that 
\begin{equation}
\frac{\left|k_{P1},...k_{PN}\right\rangle }{\left|k_{1},...k_{N}\right\rangle }=\prod_{i,j\in P}S^{-1}\left(k_{i},k_{j}\right)=\prod_{i,j\in P}S^{*}\left(k_{i},k_{j}\right)\label{eq:Eigenstate_ratio}
\end{equation}
which may be derived by considering the ratio of the components of
$\left|k_{P_{1}}...k_{P_{N}}\right)$ of the two states $\left|k_{1}...k_{N}\right\rangle $
and $\left|k_{P_{1}}...k_{P_{N}}\right\rangle $. Now in order to
prove that the two resolutions of unity in Eqs. (\ref{eq:resolutionofidentityfinitesizefinal-1-1})
and (\ref{eq:finitesizeresolutionofidentity-1}) are equal we need
to show that: 
\begin{equation}
\left|k_{n_{1}},...k_{n_{N}}\right\rangle \left\langle k_{n_{1}},....k_{n_{N}}\right|=\sum_{P\subset S_{N}}\left|k_{P_{1}}...k_{P_{N}}\right\rangle \left(k_{P_{1}}...k_{P_{N}}\right|\label{eq:ket_equality}
\end{equation}
in the quadrant $x_{1}<....<x_{N}$. However this follows directly
from Eqs. (\ref{eq:Eigenstates_quadrant}) and (\ref{eq:Eigenstate_ratio})
and the resolution of unity follows. 

\begin{thebibliography}{10}
\bibitem[22]{key-33-1} There is a similar formula for the expectation
value of several operators e.g. $\left\langle \Theta_{1}\left(t_{1}\right)\Theta_{2}\left(t_{2}\right)....\Theta_{n}\left(t_{n}\right)\right\rangle $
involving multiple resolutions of identity.
\bibitem[34]{key-32-1}$B\left(k_{i}\right)$ are Algebraic Bethe
Ansatz - ABA - operators, or matrix elements of the ABA transfer matrix.
Recall that we may write that the transfer matrix between $-L/2$
and $L/2$ in the form $T\left(k\right)\equiv$ 
\begin{equation}
\left(\begin{array}{cc}
A\left(k\right) & B\left(k\right)\\
C\left(k\right) & D\left(k\right)
\end{array}\right)=L\left(n\mid k\right)L\left(n-1\mid k\right)...L\left(1\mid k\right)
\end{equation}
Here $L\left(m\mid k\right)=$: 
\begin{equation}
\left(\begin{array}{cc}
1-i\frac{k\Delta}{2} & -i\sqrt{c}b^{\dagger}\left(m\Delta-L/2\right)\Delta\\
i\sqrt{c}b\left(m\Delta-L/2\right)\Delta & 1+i\frac{k\Delta}{2}
\end{array}\right),\label{eq:Transition_Operator-1}
\end{equation}
with $\Delta=L/n$ and take the limit $n\rightarrow\infty$. Using
this we can show that the normalization ratio between the two types
of eigenstates is given by: 
\begin{equation}
\frac{B\left(k_{1}\right)...B\left(k_{N}\right)\left|0\right\rangle }{\left|k_{1},....k_{N}\right\rangle }=\frac{\left(-i\sqrt{c}\right)^{N}\prod_{i<j}\left(k_{j}-k_{i}-ic\right)}{\prod_{i<j}\left(k_{j}-k_{i}\right)}
\end{equation}
\bibitem[35]{key-34-1} This is also a feature of all higher order
terms in the expansion.
\end{thebibliography}
\end{document}